\documentclass[11pt]{article}
\usepackage{graphics}

\makeatletter

\usepackage{aaspp4}
\renewcommand{\maketitle}{}
\newcommand{\bmath}[1]{\mbox{\boldmath$#1$}}

\makeatother

\begin{document}

\title{Enhanced Microlensing by Stars Around the Black Hole in the Galactic Center}

\author{Tal Alexander}

\maketitle
\affil{Space Telescope Science Institute, 3700 San Martin Drive, Baltimore, MD 21218}

\author{\and Abraham Loeb}

\affil{Harvard-Smithsonian Center for Astrophysics, 60 Garden Street,Cambridge, MA 02138}

The effect of stars on the lensing properties of the supermassive black
hole in the Galactic Center is similar to the effect of planets on microlensing
by a star. We show that the dense stellar cluster around SgrA$ ^{\star } $
increases by factors of a few the probability of high-magnification lensing
events of a distant background source by the black hole. Conversely, the
gravitational shear of the black hole changes and enhances the microlensing
properties of the individual stars. The effect is largest when the source
image lies near the Einstein radius of the black hole ($ 1.75^{\prime \prime }\pm 0.20'' $
for a source at infinity). We estimate that the probability of observing
at least one distant background star which is magnified by a factor $ >5 $
in any infrared snapshot of the inner $ \sim  $2{}'' of the Galactic
Center is $ \sim 1\% $ with a $ K $-band detection threshold of $ 20 $
mag. The largest source of uncertainty in this estimate is the luminosity
function of the background stars. The gravitational shear of the black
hole lengthens the duration of high-magnification events near the Einstein
radius up to a few months, and introduces a large variety of lightcurve
shapes that are different from those of isolated microlenses. Identification
of such events by image subtraction can be used to probe the mass function,
density and velocity distributions of faint stars near the black hole,
which are not detectable otherwise.

\keywords{Galaxy: center --- gravitational lensing --- infrared: stars ---Galaxy: stellar content}

\section{Introduction}

\label{sec: intro}

Deep infrared observations of the innermost region around the massive black
hole (BH) in the Galactic Center (GC) reveal numerous point sources (Genzel
et al. \cite{Gen97}; Ghez et al. \cite{Ghe98}). Most of these are probably
stars orbiting deep in the potential of the BH. However, the BH will also
gravitationally lens and magnify any background star that happens to lie
behind it, and so a small fraction of these point sources could be lensed
images of distant background stars. The possible existence of such images
in the innermost GC, beyond being a probe of the BH potential, may also
affect the apparent star counts, radial stellar density distribution and
infrared luminosity function. Previous investigations of gravitational
lensing by the BH (Wardle \& Yusef-Zadeh \cite{War92}; Alexander \& Sternberg
\cite{Ale99b}) suggested that lensing effects do not play a major role
in present-day observations of the GC. However, the BH is surrounded by
a very dense stellar cluster (e.g. Genzel et al. \cite{Gen97}), whose
contribution to the lensing by the BH has so far been neglected. Because
the stellar mass is not smoothly distributed around the BH but is composed
of discrete point masses, its effect on the lensing properties of the BH
is much larger than one may naively estimate by adding the stellar mass
to that of the BH.

The effect of a star on lensing by the BH is analogous to the effect of
a planet on microlensing by its star. The latter problem has been studied
in great detail in the context of microlensing searches for planets (Mao
\& Paczynski \cite{Mao91}; Gould \& Loeb \cite{Gou92}; Bolatto \& Falco
\cite{Bol94}; Gaudi \& Gould \cite{Gau97}; Wambsganss 1997; Peale \cite{Pea97};
Griest \& Safizadeh \cite{Gri98}; Gaudi, Naber, \& Sackett \cite{Gau98};
Gaudi \& Sackett \cite{Gau00}). In particular, Gould \& Loeb (\cite{Gou92})
first showed that the cross-section for magnification by the planet can
be increased by up to an order of magnitude due to the gravitational shear
of the star; the effect being most pronounced when the planet lies near
the Einstein radius of the star. Aside from the change in scales, the same
result should apply to SgrA$ ^{\star } $. In the GC, the primary massive
lens is the BH and the secondary low-mass lenses are the stars around it.
If a star happens to pass near the image of a source that is being lensed
by the BH, then the magnification of the image could be enhanced significantly.
The BH shear increases the probability for high magnification events by
individual stars. Miralda-Escud\'{e} \& Gould (\cite{Mir00}) pointed
out that the existence of a cluster of stellar mass BHs around the massive
BH could be detected by this effect. In this paper we consider normal stellar
lenses and calculate the probability of high-magnification events in the
stars-BH system in comparison to the naked BH case.

The statistics of microlensing by stars in the GC region is very different
from that in the Galactic disk or halo. In the latter case, the optical
depth for microlensing is very small, $ \tau \sim 10^{-7} $--$ 10^{-5} $,
but there are many background sources (see, e.g. review by Paczynski \cite{Pac96}).
In contrast, the optical depth for microlensing by stars within the inner
arcsecond of the GC is as high as $ \sim \! 0.1 $ due to the high stellar
density there (see \S\ref{sec:results}), but it is not clear whether there
is a sufficient number of bright background sources. In fact, due to their
large distances and high extinction by dust, most sources behind the GC
will only be observable while being magnified during a lensing event. The
microlensing events are transient; any distant source behind the GC will
be microlensed for a fraction of the time due to foreground stars which
are passing in front of its images.

This paper is organized as follows. The method of our calculation is described
in \S\ref{sec:method}; the models of the stellar lenses and background
stellar sources are described in \S\ref{sec:model}; and the numerical
results are presented in \S\ref{sec:results}. We discuss the main implications
of our results in \S\ref{sec:discuss}.

\section{Method}

\label{sec:method}

\begin{figure}
{\centering \resizebox*{!}{0.5\textheight}{\includegraphics{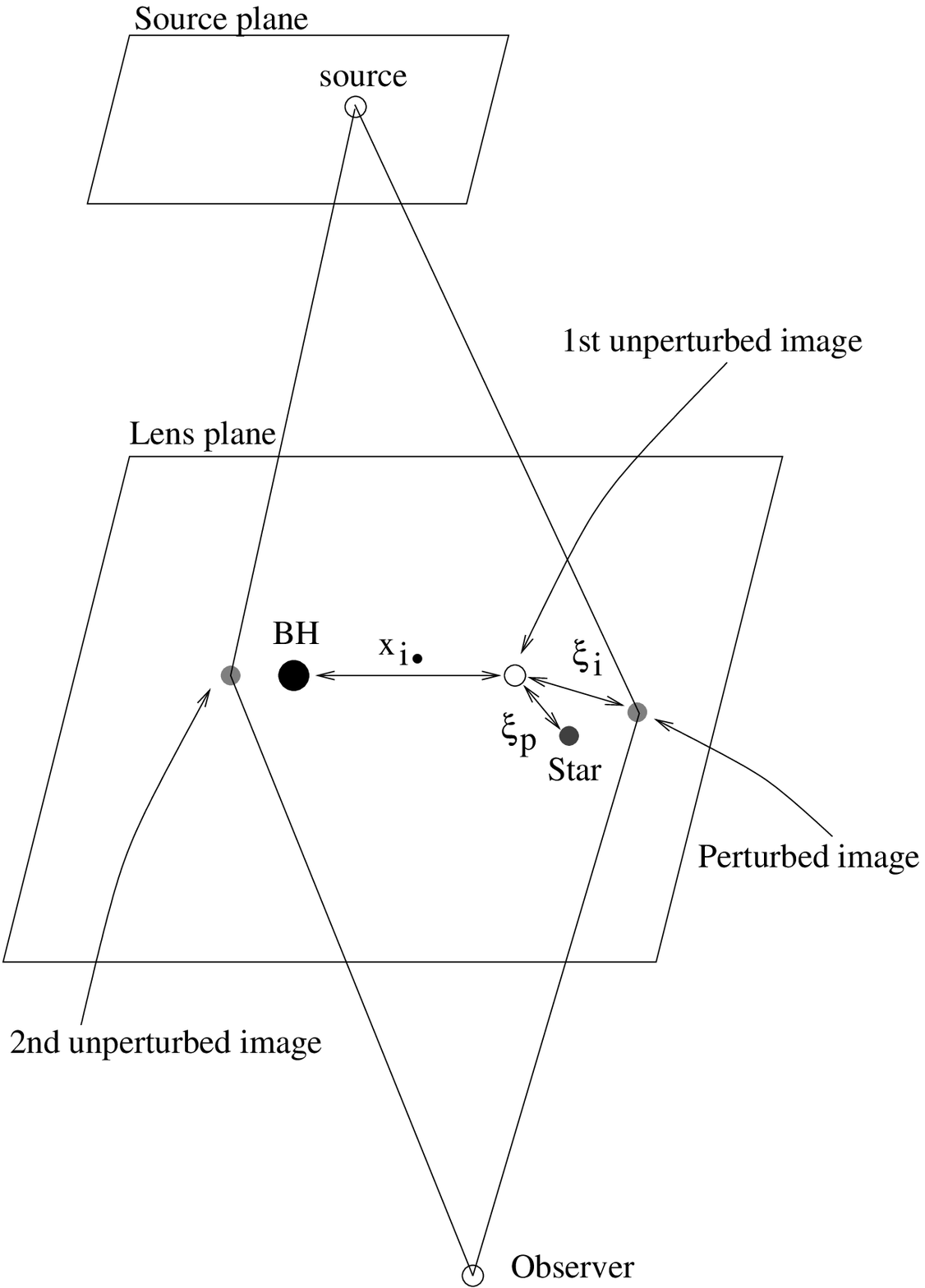}} \par}

\caption{\label{fig:coords} A sketch defining the notation used in this paper.
The presence of a perturbing star at position $\bmath {\xi}_{p}$
relative to the unperturbed image at $\bmath {x}_{i\bullet}$ splits
this image into 2 or 4 images at positions $\bmath {\xi}_i$ relative
to the unperturbed image. For clarity, only one of the multiple images
due the the star is shown and the proportions are exaggerated.}
\end{figure}

We adopt the semi-analytic method of Gould \& Loeb (\cite{Gou92}; see
also the original discussion by Chang \& Refsdal \cite{Cha79}, \cite{Cha84})
to describe the effect of a low-mass secondary lens (a star) on the image
of a point source produced by a high-mass primary lens (the BH). The limitations
of the point source assumption are discussed in \S\ref{sec:results}. We
consider rare, high-magnification events for which the possibility that
more than one star affects the microlensing event at any given time can
be neglected. The original, unperturbed image position at $ \bmath {x}_{i\bullet } $
is displaced by $ \bmath {\xi }_{i}=(\xi _{i},\eta _{i} $) due to the
perturbing star at a position $ \bmath {\xi }_{p}=(\xi _{p},\eta _{p}) $
relative to the unperturbed image (Fig. \ref{fig:coords}). We express
the source position $ \bmath {x}_{s} $ and the unperturbed image position
$ \bmath {x}_{i\bullet } $ in terms of the Einstein angular radius of
the isolated BH, \begin{equation}
\theta _{E}\equiv \left[ {4GM_{\bullet }\over c^{2}}{(D_{s}-R_{0})\over D_{s}R_{0}}\right] ^{1/2}=\left( 1-{R_{0}\over D_{s}}\right) ^{1/2}\theta _{\infty },
\end{equation}
 where $ M_{\bullet }=(3.0\pm 0.5)\times 10^{6}\, M_{\odot } $ (Genzel
et al. \cite{Gen00}) is the BH mass, $ R_{0}=8.0\pm 0.5\, \mathrm{kpc} $
(Reid \cite{Rei93}) is the distance to the GC, $ D_{s} $ is the distance
to the source, and $ \theta _{\infty }=1.75^{\prime \prime }\pm 0.20^{\prime \prime } $
is the Einstein angle for a source at infinity ($ 1''\simeq 0.04\textrm{ pc} $
at the GC). We express the displacement vectors $ \bmath {\xi }_{i} $
and $ \bmath {\xi }_{p} $ in terms of the Einstein radius of the star,
$ \sqrt{\epsilon }\theta _{E} $, where $ \epsilon =m_{\star }/M_{\bullet } $
is the mass ratio between the star and the BH. This quantity provides a
good measure for the scale of the stellar lensing zone. Correspondingly,
we express areas in the lens plane, $ \sigma _{\star } $, in units of
$ \epsilon \pi \theta _{E}^{2} $, the Einstein ring area of an isolated
star; and express surface number densities of lensing stars in the lens
plane, $ \Sigma _{\star } $, by $ \left( \epsilon \pi \theta _{E}^{2}\right) ^{-1} $.
Areas in the source plane, $ \sigma _{s} $, are expressed in terms of
$ \pi \theta _{E}^{2} $, and surface number density of sources in the
source plane , $ \Sigma _{s} $, are correspondingly expressed in terms
of $ \left( \pi \theta _{E}^{2}\right) ^{-1} $.

In the limit $ \epsilon \ll 1 $, $ \xi _{i} $ is given by the two
or four solutions to the equations (Gould \& Loeb \cite{Gou92}) \begin{eqnarray}
\xi _{i}^{4}+\frac{(1-2\gamma )\xi _{p}}{\gamma }\xi ^{3}_{i}+\left[ \frac{(1-\gamma )^{2}(\xi ^{2}_{p}+\eta ^{2}_{p})}{4\gamma ^{2}}-\frac{(1-\gamma )\xi _{p}^{2}}{\gamma }-\frac{1}{1+\gamma }\right] \xi _{i}^{2} & - & \nonumber \\
\left[ \frac{(1-\gamma )^{2}(\xi ^{2}_{p}+\eta ^{2}_{p})\xi _{p}}{4\gamma ^{2}}+\frac{(1-\gamma )\xi _{p}}{\gamma (1+\gamma )}\right] \xi _{i}-\frac{(1-\gamma )^{2}\xi _{p}^{2}}{4\gamma ^{2}(1+\gamma )} & = & 0\, ,\label{eq:xi} 
\end{eqnarray}
 where $ \gamma \equiv x^{-2}_{i\bullet } $, and \begin{equation}
\eta _{i}=\frac{(1+\gamma )\eta _{p}\xi _{i}}{2\gamma \xi _{i}+(1-\gamma )\xi _{p}}\, .
\end{equation}
 The magnification of each of the images is \begin{equation}
\label{eq:A}
A=\left| 1-\left[ \gamma +(1+\gamma )^{2}\xi ^{2}_{i}-(1-\gamma )^{2}\eta ^{2}_{i}\right] ^{2}-4(1-\gamma ^{2})^{2}\xi ^{2}_{i}\eta ^{2}_{i}\right| ^{-1}\, .
\end{equation}
 For the GC, we consider solar mass lenses and adopt, for simplicity, a
single value of $ \epsilon =3\times 10^{-7} $.

The angular area in the source plane where a background source would be
magnified above a given threshold $ A $ by the star-BH system, $ \sigma _{s}(>\! A) $,
is obtained by identifying the corresponding angular area in the lens plane
around the unperturbed image and transforming it back to the source plane\footnote{%
Source areas that contribute two images which are above the magnification
threshold, are counted twice. 
} . The lensing stars are distributed randomly around the BH and they scan
the lens plane as they move. Averaging over time and assuming an isotropic
projected stellar distribution around the BH, \begin{equation}
\label{eq:ss}
\sigma _{s}(>\! A)=\frac{1}{\pi }\int P({>A},{\bmath {x}_{i\bullet }})\left| \frac{d\bmath {x}_{s}}{d\bmath {x}_{i\bullet }}\right| d\bmath {x}_{i\bullet }=2\int P({>A},{\bmath {x}_{i\bullet }})A^{-1}_{\bullet }(x_{i\bullet })x_{i\bullet }dx_{i\bullet }\, ,
\end{equation}
 where $ P({>A},{\bmath {x}_{i\bullet }}) $ is the probability for magnification
above $ A $, that is, the fraction of $ d\bmath {x}_{i\bullet } $
where the image is magnified above $ A $. Equivalently, $ P({>A},{\bmath {x}_{i\bullet }}) $
is the fraction of time a stationary source is magnified above $ A $
by a closely-passing star. The magnification $ A_{\bullet } $ is that
due to the BH alone, \begin{equation}
A_{\bullet }=\left| {d\bmath {x}_{i\bullet }\over d\bmath {x}_{s}}\right| =\left| 1-{x^{-4}_{i\bullet }}\right| ^{-1},
\end{equation}
 where the positions of these images are related to the source position,

\begin{equation}
\label{eq:Abh}
x_{i\bullet ,\pm }=\frac{1}{2}\left( x_{s}\pm \sqrt{x_{s}^{2}+4}\right) \, .
\end{equation}
 De-magnification, i.e. $ A<A_{\bullet } $, is also possible. The probability
$ P $ is expressed in terms of the optical depth \begin{equation}
\label{eq:tau}
\tau _{\star }\left( >\! A,x_{i\bullet }\right) \equiv \Sigma _{\star }(x_{i\bullet })\sigma _{\star }(>\! A,x_{i\bullet })\, ,
\end{equation}
 where $ \sigma _{\star }(>\! A,x_{i\bullet }) $ is the area in the
lens plane where the presence of a star will lead to magnification above
$ A $, and $ \Sigma _{\star }(x_{i\bullet }) $ is the stellar lens
surface density. In the small optical depth limit \begin{equation}
\label{eq:PA}
P({>A},{\bmath {x}_{i\bullet }})\simeq \tau _{\star }(>\! A,x_{i\bullet })+\Theta \left( A_{\bullet }-A\right) \, ,
\end{equation}
 where $ \Theta  $ is the Heaviside step function. In practice, the
lens plane area $ \sigma _{\star }(>\! A,x_{i\bullet }) $ is calculated
by solving equations (\ref{eq:xi})--(\ref{eq:A}) numerically as functions
of $ \bmath {\xi }_{p} $ in a small test area $ \sigma _{0}(x_{i\bullet }) $
around $ \bmath {x}_{i\bullet } $, where the effect of the stellar lens
magnification is appreciable. The effect of the star outside of $ \sigma _{0} $
is neglected. With this assumption, the probability for not being affected
by any star is $ \exp (-\tau _{0}) $, where $ \tau _{0}\equiv \Sigma _{\star }\sigma _{0} $.
The magnification probability is then given by the sum of the probabilities
of magnification by a star or by the BH alone, $ P(>\! A)=\left[ 1-\exp (-\tau _{\star })\right] +\exp (-\tau _{0})\Theta (A_{\bullet }-A) $,
which in the limit $ \tau _{\star },\, \tau _{0}\ll 1 $ reduces to equation
(\ref{eq:PA}), independently of the exact choice of $ \sigma _{0} $.

For an isolated BH ($ \Sigma _{\star }=0 $), the corresponding angular
area in the high magnification limit, $ A\gg 1 $, is $ \sigma _{s}(>\! A)_{\bullet }\approx \! 1\left/ 2A^{2}\right.  $.

Equations (\ref{eq:xi})--(\ref{eq:A}) are valid as long as $ \tau _{\star }(>\! A,x_{i\bullet })\ll 1 $,
i.e. when $ A $ is large enough so that there is no overlap between
the regions of influence (lensing zones) of different stars, $ \sigma _{\star }(>\! A,x_{i\bullet })\ll \Sigma ^{-1}_{\star }(x_{i\bullet }) $.
In addition, these equations ignore the cumulative effect of all the stars
on the central caustic around the BH (Griest \& Safizadeh 1998). For $ \epsilon =3\times 10^{-7} $,
the central caustic has a negligible contribution to the integrand of equation
(\ref{eq:ss}) as long as $ x_{i\bullet } $ is not very close to unity
(i.e. for sources which are not located almost behind the BH). The inclusion
of the central caustic can only increase our estimated lensing probabilities.

\section{Model}

\label{sec:model}

In \S\ref{sec:results} below we demonstrate the effect of enhanced microlensing
by stars near a massive BH by calculating the mean number of lensed images
with magnification above a threshold $ A $, $ \left\langle N_{i}(>\! A)\right\rangle =\sigma _{s}(>\! A)\Sigma _{s} $,
of distant background sources behind SgrA$ ^{\star } $. To do this,
we need to specify the projected density of the stellar lenses $ \Sigma _{\star } $,
which determines the optical depth for microlensing and hence the cross-section
in the source plane $ \sigma _{s}(>\! A) $ (eq. {[}\ref{eq:ss}{]}),
and we also need to specify the projected source density $ \Sigma _{s} $
and source luminosity function. We express these surface densities scaled
to $ \theta _{\infty }=1.75'' $.

\subsection{Stellar Lenses Around SgrA$ ^{\star }  $}

The stellar mass density distribution near the BH is modeled as a power-law,
\begin{equation}
\rho =(3-\alpha )\rho _{b}\left( {r\over r_{b}}\right) ^{-\alpha },
\end{equation}
 with $ \rho _{b}=10^{6}\, M_{\odot }\, \mathrm{pc}^{-3} $, $ r_{b}=0.4\, \mathrm{pc} $
and an index $ \alpha  $ in the range of $ 3/2 $ to $ 7/4 $. The
normalization of the mass density is based on the dynamic mass measurements
of Eckart \& Genzel (\cite{Eck97}). The power-law distribution is indicated
by an analysis of the star counts in the inner GC (Alexander \cite{Ale99a})
and agrees with the theoretical prediction for a relaxed stellar system
around a black hole (Bahcall \& Wolf \cite{Bah77}), as is thought to be
the case in the GC. For simplicity, we assume that the stellar lenses all
have the same mass $ m_{\star }=1\, M_{\odot } $ ($ \epsilon \approx 3\times 10^{-7} $).
For these low mass stars, we adopt the theoretical prediction of $ \alpha \sim \! 3/2 $.
The corresponding surface mass density (in units of $ M_{\bullet }\left/ \pi \theta ^{2}_{E}\right.  $)
and surface number density (in units of $ 1\left/ \epsilon \pi \theta ^{2}_{E}\right.  $)
at an angular separation $ x=\theta /\theta _{E} $ from the GC are both
given by \begin{equation}
\label{eq:Ss}
\Sigma _{\star }=\widehat{\Sigma }_{\star }x^{1-\alpha }\, ,
\end{equation}
with $ \widehat{\Sigma }_{\star }\simeq 0.04\left( \left. \theta _{E}\right/ \theta _{\infty }\right) ^{3-\alpha } $.
The normalization $ \widehat{\Sigma }_{\star } $ is almost independent
of $ \alpha  $ in the above range and scales as $ M^{-1}_{\bullet } $.
The ratio of stellar mass enclosed within the Einstein radius and the mass
of the BH is $ 2\widehat{\Sigma }_{\star }\left/ (3-\alpha )\right.  $.
Note that $ \Sigma _{\star } $ is also the optical depth for lensing
by the stars alone, neglecting the effect of the BH. As we show below,
the actual optical depth increases substantially when the shear of the
BH is included, an effect analogous to the cross-section enhancement for
planetary systems (Gould \& Loeb 1992).

\subsection{Distant Background Stars}

A rough estimate of the projected density of distant stars, which is based
on a model of the $ K $-band luminosity distribution in the Galaxy (Kent
\cite{Ken92}) and the assumption that the mean stellar $ K $-band luminosity
is solar, suggests that there are of order $ \sim \! 100 $ distant background
stars within $ \sim \! 2^{\prime \prime } $ of the BH. We define {}``distant
background stars{}'' as those stars with $ \theta _{E}\gtrsim 1^{\prime \prime } $,
i.e. farther than $ \sim \! 4 $ kpc behind the BH. The surface density
of nearby background stars in the inner bulge and in the high density central
cluster may be as high as $ \sim \! 10^{3}\, \mathrm{arcsec}^{-2} $
(Alexander \& Sternberg \cite{Ale99b}). However, the Einstein angular
radius for stars so close behind the BH is less than $ 0.1'' $ and there
are not enough lensing stars near the BH on that scale for microlensing
to become important. Here, we study the role of these close stars as microlenses
rather than as sources.

The stellar $ K $ luminosity function (KLF) of the distant background
stars can be determined from observations in the $ b=0^{\circ } $, $ l=30^{\circ } $
direction (i.e. impact parameter of 4 kpc relative to the GC). The
infrared Galactic model of Ortiz \& L\'{e}pine (\cite{Ort93}), which
reproduces such observations, predicts $ \Sigma _{s}(<\! 13.5^{\mathrm{m}})\sim \! 0.1 $
(see their Fig. 17b) with a steep slope of $ b\equiv d\log \Sigma _{s}/dK\sim \! 0.4 $
to $ 0.5 $. The projected stellar population is integrated over most
of the Galactic disk and is composed of stars of different ages, and so
it is reasonable to approximate it as an old, continuously star forming
population. Observations of such populations in the Galactic Bulge (Tiede
et al. \cite{Tie95}; Holtzman et al. \cite{Hol98}) and the central cluster
in the GC (Blum, Sellgren \& DePoy \cite{Blu96}; Davidge et al. \cite{Dav97}),
as well as stellar population synthesis of the GC (Alexander \& Sternberg
\cite{Ale99b}), indicate that the post-main-sequence KLF, where the $ K $-luminous
sources lie, typically takes the form of a single power-law with $ b\sim 0.3 $--$ 0.4 $
down to the main sequence turn-off point at around $ M_{K\odot }=3.3^{\mathrm{m}} $
, where the power-law flattens and then turns over. Assuming a typical
distance of $ 2R_{0} $ to the distant background sources and $ A_{K}\sim 3^{\mathrm{m}} $
to the GC (Blum et al. \cite{Blu96}), and using the population synthesis
models to estimate the mean stellar $ K $ luminosity and the fraction
of stars less massive than solar, we approximate the KLF as a single power
law that extends down to a cutoff magnitude of $ K_{c}\sim \! 28.2 $
mag and normalize it by Kent's integrated luminosity density model. The
uncertainty in the exact shape of the luminosity function, the patchiness
of the extinction in the Galactic Plane, and the distribution of star-forming
regions in the spiral arms behind the GC introduce a large uncertainty
to $ \Sigma _{s} $. Putting these uncertainties aside, we adopt a working
model for the KLF of the distant background stars, \begin{equation}
\label{eq:KLF}
\Sigma _{s}(<K)=\widehat{\Sigma }_{s}10^{bK}\, ,
\end{equation}
 with $ \widehat{\Sigma }_{s}=5\times 10^{-10}\left( \left. \theta _{E}\right/ \theta _{\infty }\right) ^{2} $
and $ b=0.4 $. 

The uncertainties inherent in the estimate of $ \widehat{\Sigma }_{s} $
are demonstrated by the attempt to extrapolate to low luminosity the observed
KLF and correct it for distance, projection, foreground stars and the higher
extinction at $ l=0^{\circ } $. The Galactic extinction model of Wainscoat
et al. (\cite{Wai92}) predicts that the integrated $ K $-band extinction
in the direction of the GC is $ \sim 2^{\mathrm{m}} $ larger than that
at $ l=30^{\circ } $. The correction for foreground stars and for the
projection from $ l=30^{\circ } $ to $ l=0^{\circ } $ (assuming an
exponential Galactic disk with a length scale of 3.5 kpc, Wainscoat et
al. \cite{Wai92}) further reduces the projected density by a factor of
$ \sim \! 0.5 $, and the shift from a typical distance of $ \sim \! R_{0} $
to $ 2R_{0} $ by another factor of $ \sim \! 0.25 $. This gives $ \Sigma _{s}(<15.5^{\mathrm{m}})\sim \! 0.01\left( \left. \theta _{E}\right/ \theta _{\infty }\right) ^{2} $
for the distant background stars, which corresponds to $ \widehat{\Sigma }_{s}=5\times 10^{-9}\left( \left. \theta _{E}\right/ \theta _{\infty }\right) ^{2} $,
10 times higher than our adopted normalization (eq. \ref{eq:KLF}).

The differential surface density of lensed stars, $ d{\widetilde{\Sigma }_{s}}/d{\widetilde{F}}|_{\widetilde{F}} $,
is related to their unlensed luminosity function by $ d{\widetilde{\Sigma }_{s}}/d{\widetilde{F}}|_{\widetilde{F}}=A^{-2}d{\Sigma _{s}}/d{F}|_{\widetilde{F}/A} $,
where the tilde denotes lensed quantities and where $ d{\Sigma _{s}}/d{F}|_{F}\propto F^{-2.5b-1}\propto F^{-2} $
for the KLF of equation (\ref{eq:KLF}). Hence, the adopted surface density
has the critical power-law slope for which the differential flux distribution
remains unchanged.

\section{Results }

\label{sec:results}

\begin{figure}
\begin{tabular}{c}
 \resizebox*{!}{0.4\textheight}{\includegraphics{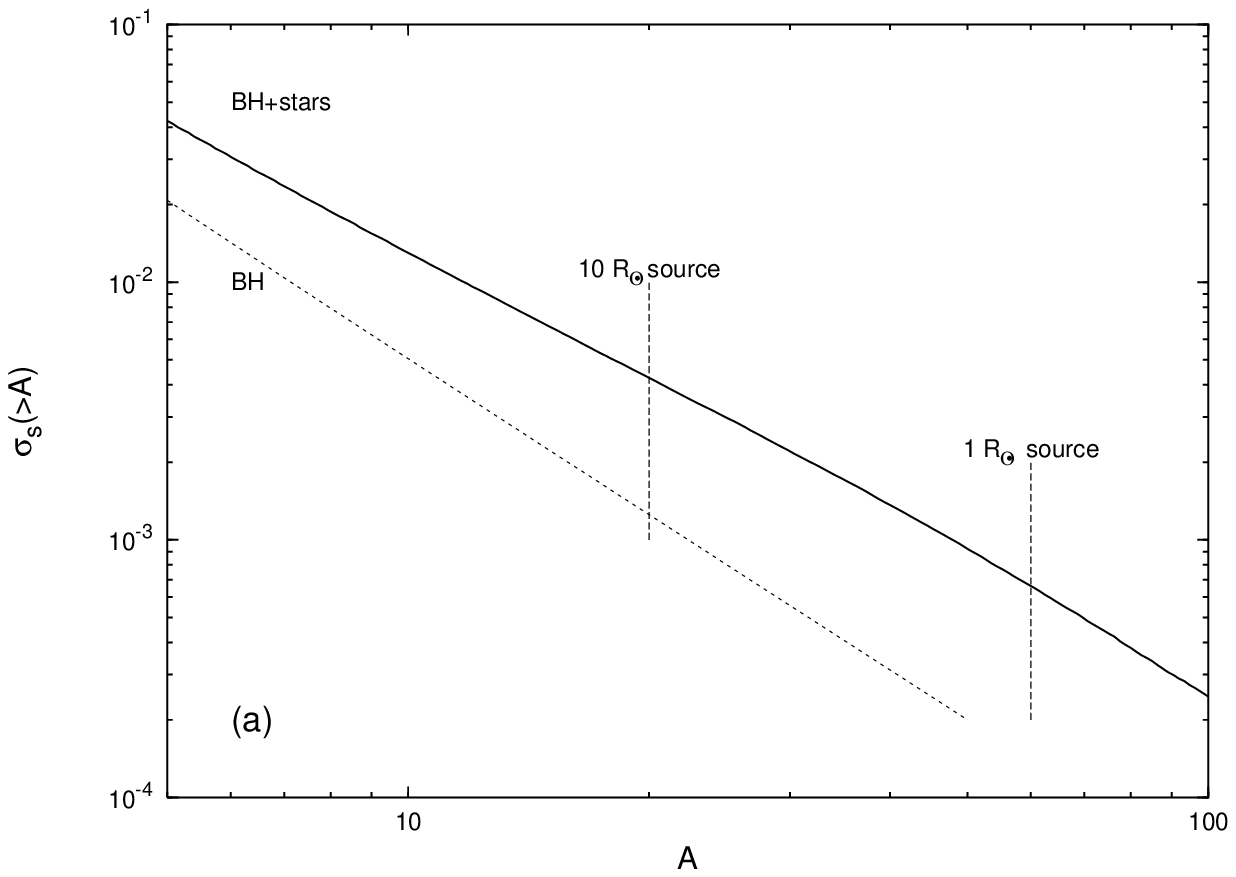}}\\
 \resizebox*{!}{0.4\textheight}{\includegraphics{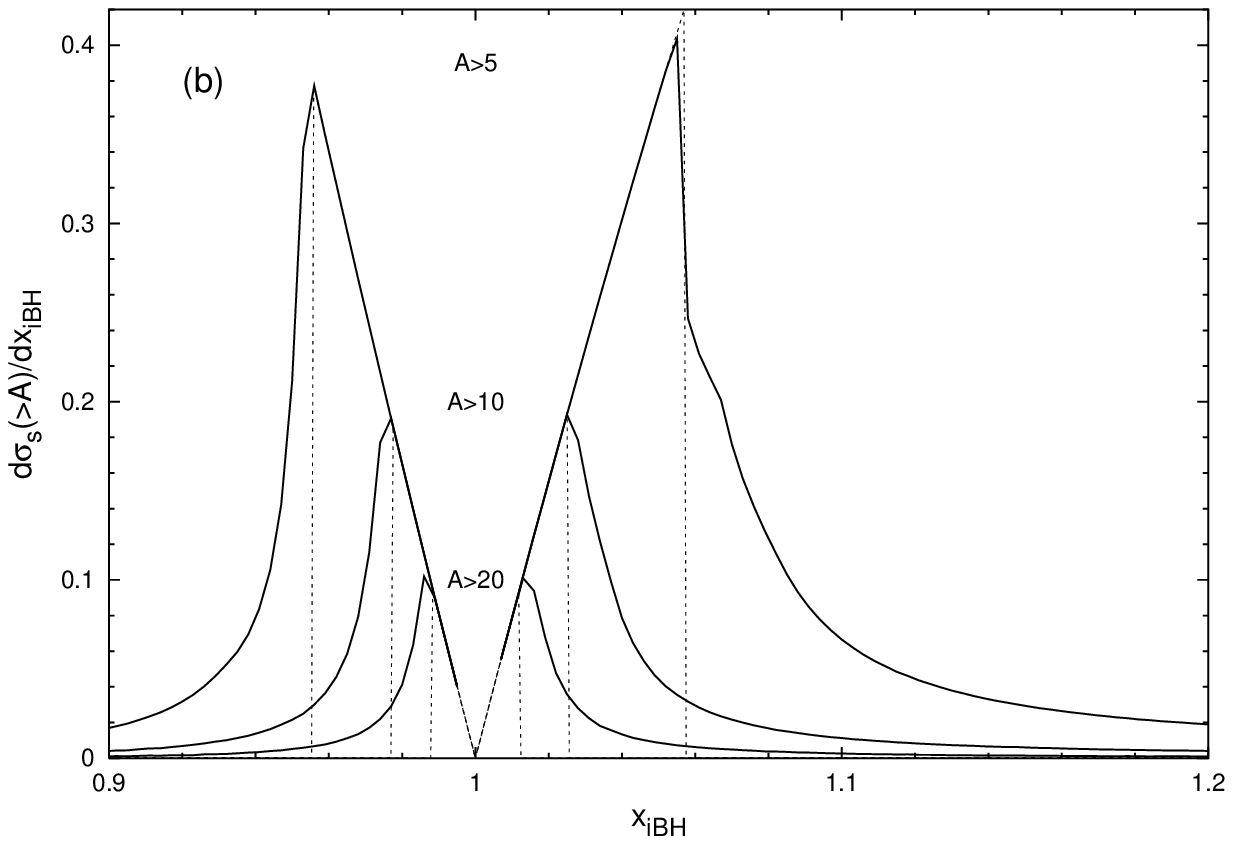}} \\
\end{tabular}

\caption{\label{fig:apdf}(a) The mean area in the source plane for magnification
by more than $A$ by the BH alone, and by the BH and stars of an image
in the inner $2\theta_{E}$, for $\theta_{E}=\theta_{\infty}$ and the
stellar lens surface density model of equation (\protect\ref{eq:Ss})
with $\alpha =3/2$ (see text for scaling to arbitrary $\theta _{E}$).
The two vertical hash marks roughly indicate the maximal magnification
possible for extended sources, a $1\, R_{\odot}$ and a $10\,
R_{\odot}$ star at a distance of $2R_{0}$. (b) The differential source
area per annulus in the lens plane that is magnified by more than
$A$. The case of a BH and stars (full lines) is compared to the
contribution from the BH alone (dashed lines) for three threshold
values of $A$.}
\end{figure}

Figure \ref{fig:apdf}a shows the integrated source plane area $ \sigma _{s}(>\! A) $
for producing a lensed image of a point source that is magnified by more
than $ A $ anywhere in the inner $ 2\theta _{E} $, for $ \theta _{E}=\theta _{\infty } $
and the stellar lens surface density model of equation (\ref{eq:Ss}) with
$ \alpha =3/2 $. The results can be scaled to any value of $ \theta _{E} $
or any stellar lens surface density normalization $ \widehat{\Sigma }_{\star } $
by noting that $ \sigma _{s}(>\! A)-\sigma _{s}(>\! A)_{\bullet }\propto \widehat{\Sigma }_{\star }\propto \theta ^{3-\alpha }_{E} $
(Eqs. {[}\ref{eq:ss}{]}, {[}\ref{eq:PA}{]} and {[}\ref{eq:Ss}{]}).
In particular, for $ D_{s}=2R_{0} $ the difference in source plane areas
is decreased by a factor of 0.6. The highest magnifications are due to
stellar lenses close to $ \theta _{E} $. The integration range in equation
(\ref{eq:ss}) does not include the annulus $ \left| 1-x_{i\bullet }\right| <\delta x=0.005 $.
At that point the semi-analytic approximation (eq. {[}\ref{eq:xi}{]})
and the small optical depth assumption no longer apply because the change
in $ A_{\bullet } $ over the angular scale defined by $ \sigma _{\star }(>\! A,x_{i\bullet }) $
is no longer small, and $ \tau _{\star }(>\! A,x_{i\bullet })\ll 1 $
no longer holds even for very high values of $ A $. We verified the
validity of our semi-analytic calculations for the case of a BH and a single
star in the range $ \delta x>0.001 $ by comparing $ \sigma _{\star }(>\! A,x_{i\bullet }) $
as function of $ A $ and $ x_{i} $, with results from the direct
ray shooting method (Wambsganss \& Kubas \cite{Wam00}). The two methods
give identical results to within $ \sim \! 5\% $ down to $ \delta x=0.1 $
for both the normalization and slope and remain in good qualitative agreement
even down to $ \delta x=0.001 $ throughout the range where the ray-shooting
method has sufficiently high resolution. 

For a star at a particular location, the magnification probability approaches
the asymptotic $ A^{-2} $ behavior in accordance with the theoretical
asymptotic limit for a point source (eq. {[}11.21b{]} of Schneider, Ehlers
\& Falco \cite{Sch92}) at some minimum magnification $ A_{b} $, which
depends on the external shear at that location. As the projected location
of the image approaches $ \theta _{E} $, the $ A^{-2} $ tail is recovered
at an increasingly larger value of $ A_{b} $. With the above choice
for $ \delta x $, we find that the area $ \sigma _{s}(>\! A) $ initially
falls off as $ A^{-3/2} $ up to a break at $ A_{b}\gtrsim 50 $ where
it begins to converge to the asymptotic $ A^{-2} $ behavior. In our
semi-analytic approach, the value of $ A_{b} $ depends on the choice
of the truncation radius $ \delta x $, with smaller values of $ \delta x $
corresponding to larger values of $ A_{b} $. An exact ray-shooting calculation
is required to determine $ A_{b} $ more precisely for a point source.
Nevertheless, we note that our calculated value for $ \sigma _{s}(>\! A) $
in the asymptotic limit agrees very well with the analytic expression for
a binary lens (eq. {[}11.23a{]} of Schneider, Ehlers \& Falco \cite{Sch92}),
taking into account the fact that our definition of $ \gamma  $ coincides
with the shear parameter in the limit $ \varepsilon \rightarrow 0 $
(Dominik \cite{Dom99}, Eqs. {[}12{]} and {[}66{]}) \begin{equation}
\sigma _{s}(>\! A\gg 1)=\left[ \frac{4}{\pi }\widehat{\Sigma }_{\star }\int x_{i\bullet }^{2-\alpha }\frac{E\left( 2\sqrt{\gamma }/\left[ 1+\gamma \right] \right) }{\left| 1-\gamma ^{2}\right| \left| 1-\gamma \right| }dx_{i\bullet }\right] \frac{1}{A^{2}}\simeq 66\frac{\widehat{\Sigma }_{\star }}{A^{2}}\, ,
\end{equation}
 where the integral was evaluated for $ \alpha =3/2 $ on the interval
$ (0\le x_{i\bullet }<1-\delta x)\cup (1+\delta x<x_{i\bullet }\le 2) $
and where $ E $ is the complete elliptic integral of the 2nd kind.

At high magnification the extended nature of the stellar sources can no
longer be neglected. The maximal magnification for an extended source of
angular size $ \theta _{s} $ is of order $ A_{\mathrm{max}}\sim \sqrt{\epsilon ^{1/2}\theta _{E}/\theta _{s}} $
(Schneider, Ehlers \& Falco \cite{Sch92}, p. 215), which for a $ 1\, R_{\odot } $
star located at $ 2R_{0} $ gives $ A_{\mathrm{max}}\sim 60 $ and
for a $ 10\, R_{\odot } $ giant at the same location gives $ A_{\mathrm{max}}\sim 20 $.
The absence of the high-magnification events is compensated by an increase
in $ \sigma _{s} $ below $ A_{\mathrm{max}} $, which is not taken
into account here. We find that even for magnifications $ A\ll A_{\mathrm{max}} $,
the cross-section of the BH and stars system is larger by factors of $ \sim 2 $
for sources at $ D_{s}=2R_{0} $, depending on the normalization of the
surface density model for the lensing stars.

Figure \ref{fig:apdf}b shows the differential area $ \mathrm{d}\sigma _{s}(>\! A)\left/ \mathrm{d}x_{i\bullet }\right.  $,
for the BH with and without the stars. The contribution of stars to the
high magnification events extends well beyond the narrow region around
the Einstein radius that is expected for an isolated BH.

\begin{figure}
\begin{tabular}{c}
   \resizebox*{!}{0.4\textheight}{\includegraphics{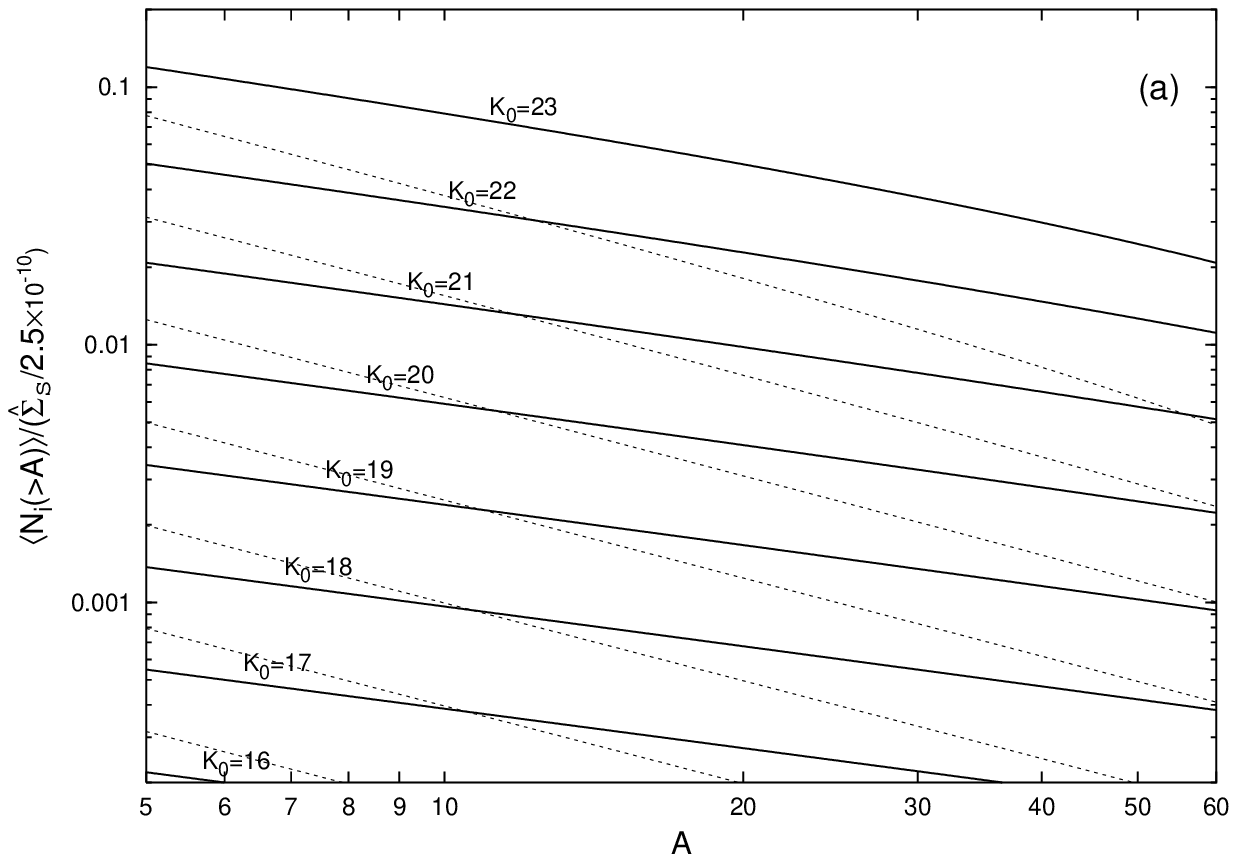}}\\
   \resizebox*{!}{0.4\textheight}{\includegraphics{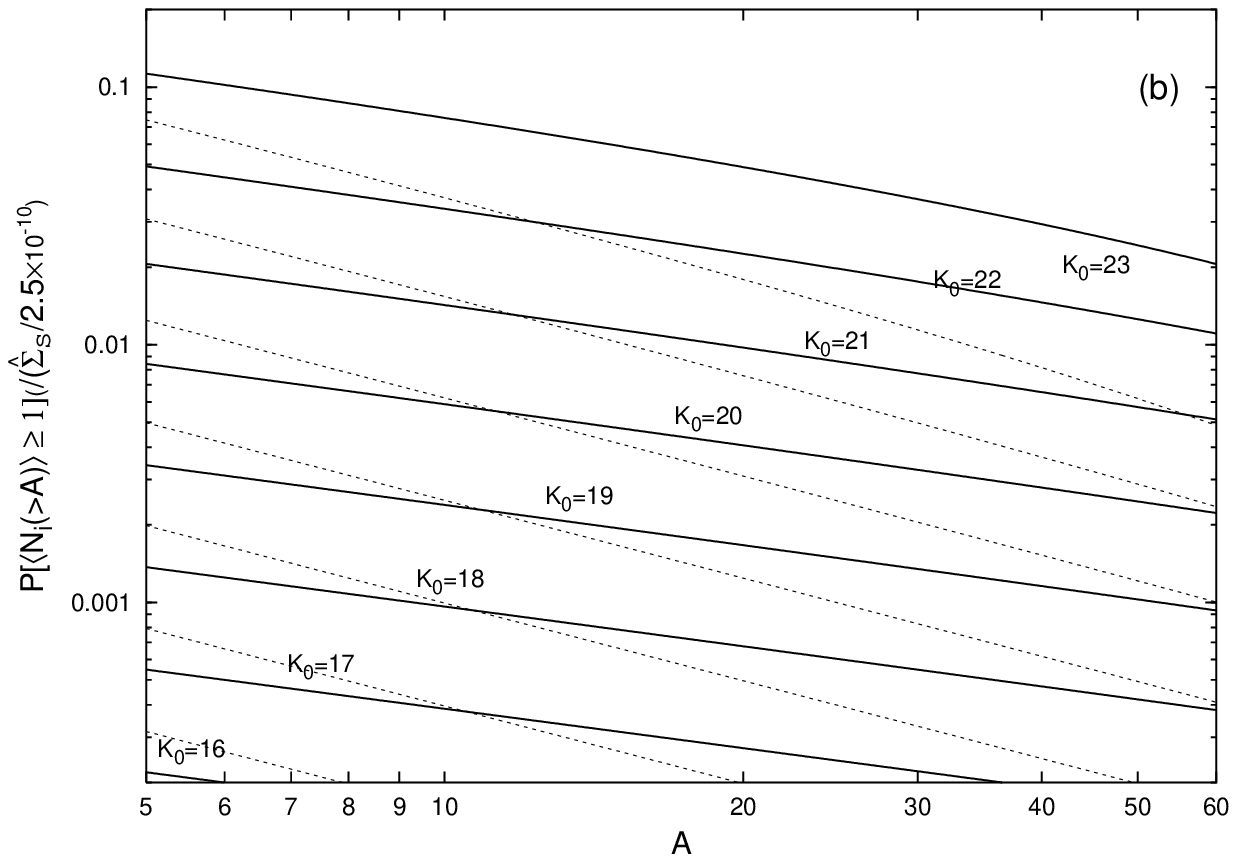}} \\
\end{tabular}

\caption{\label{fig:Ni}(a) The average number of lensed images magnified by more
than $A$ that will be observed in the inner $2\theta _{E}$ with a
limiting $K$-band magnitude $K_{0}$, for $D_{s}=2R_{0}$ and the
stellar lens surface density model of equation (\protect\ref{eq:Ss})
with $\alpha =3/2$. (b) The fraction of time that at least one lensed
image magnified by more than $ A $, will be observed in the inner
$2\theta _{E}$ with a limiting $K$-band magnitude $K_{0}$ ($P\ll 1$
scales linearly with $\widehat{\Sigma}_{s}$). The labeled solid lines
are for the BH and stars and the unlabeled dotted lines, which follow
the same $K_{0}$ order, are for the BH alone.}
\end{figure}

Figure \ref{fig:Ni}a shows the number of lensed images that will be observed,
on average, in the inner $ 2\theta _{E} $ for different limiting $ K $-magnitudes,
for $ D_{s}=2R_{0} $ and $ \alpha =3/2 $ \begin{equation}
\begin{array}{rcl}
\left\langle N_{i}(>\! A;K_{0})\right\rangle  & = & \int ^{\infty }_{A}dA'\int ^{K_{0}}_{-\infty }dK\left. \frac{d\sigma _{s}}{dA}\right| _{A'}\left. \frac{d\Sigma _{s}}{dK}\right| _{K+K_{A'}}\\
 & = & \widehat{\Sigma }_{s}\widehat{\sigma }_{s}\left\{ \begin{array}{cc}
10^{bK_{c}}A_{c}^{-1.5}+\frac{3}{5b-3}10^{bK_{0}}\left( A^{2.5b-1.5}_{c}-A^{2.5b-1.5}\right)  & \mathrm{if}\, A\leq A_{c}\\
10^{bK_{c}}A^{-1.5} & \mathrm{else}
\end{array}\right. \, ,
\end{array}
\end{equation}
 where we assume that the KLF of equation (\ref{eq:KLF}) has a sharp cutoff
beyond $ K_{c} $, and where $ K_{A}\equiv 2.5\log A $, $ A_{c}\equiv 10^{0.4(K_{c}-K_{0})} $,
and $ \sigma _{s} $ is parameterized as $ \sigma _{s}(>\! A)=\widehat{\sigma }_{s}A^{-1.5} $
with $ \widehat{\sigma }_{s}\approx 0.16 $ (Fig. \ref{fig:apdf}a).
An analogous expression with $ A^{-2} $ and a lower normalization describes
the magnification by the BH alone. For low values of $ A $ and $ K_{0} $,
$ \left\langle N_{i}(>\! A;K_{0})\right\rangle  $ decreases slowly like
$ \propto \! A^{2.5b-1.5} $ because the decrease of $ \sigma _{s}(>\! A) $
is partially offset by the ever larger number of faint stars in the population
that become accessible. However, once $ K_{0}+K_{A}>K_{c} $, all the
background stellar population can be observed when magnified by $ A $,
and $ \left\langle N_{i}(>\! A;K_{0})\right\rangle  $ falls off more
rapidly, like $ \sigma _{s}(>\! A)\propto \! A^{-1.5} $.

The motions of the stellar lenses and the sources introduce time-dependent
Poissonian fluctuations in $ N_{i}(>\! A;K_{0}) $. The fraction of time
that at least one lensed image will be observed, which is shown in Fig.
\ref{fig:Ni}b, is $ P\left[ N_{i}(>\! A;K_{0})\geq 1\right] =1-\exp \left[ -\left\langle N_{i}(>\! A;K_{0})\right\rangle \right]  $. 

Present-day observations reach a depth of $ K_{0}\sim 16^{\mathrm{m}} $
(Ghez et al. \cite{Ghe98}). For this threshold, the probability of seeing
at least one distant background star in the inner $ 2\theta _{E}\sim 2'' $
($ D_{s}=2R_{0} $) with $ A>5 $ is small, $ \sim \! 0.02\% $.
The lensing statistics will improve significantly with deeper observations.
We predict that in any snapshot of the central $ 2'' $ region around
the GC to a $ K $-magnitude limit of $ 23^{\mathrm{m}} $, at least
one distant background star will be magnified by a factor $ >5 $ for
$ 10\% $ of the time.

\begin{figure}
\par\centering \resizebox*{!}{0.4\textheight}{\includegraphics{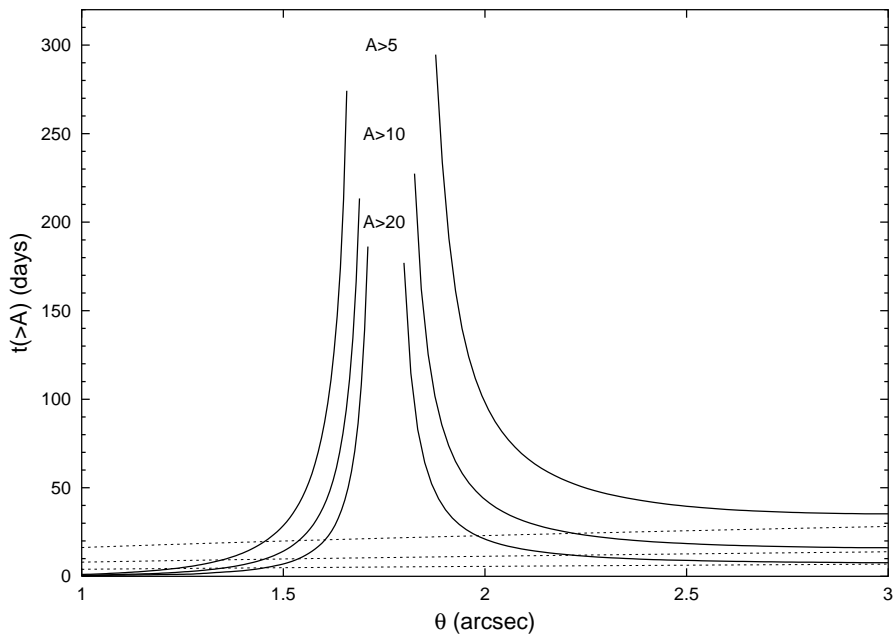}} \par{}

\caption{\label{fig:tscale}The typical duration of a microlensing event for a
stationary source at infinity by a solar mass lens (assuming the mean
rms velocity of equation{[}\protect\ref{eq:vt}{]}), as a function of
angular separation $\theta =x_{i\bullet }\theta _{\infty }$ from
SgrA$^{\star}$. Curves are shown for different magnification values
$A$, comparing stars in the shear field of the BH (full lines) with
isolated stars moving at the same velocity (dashed lines). }
\end{figure}

Variability offers a direct way of identifying the lensed images among
the many foreground stars. Any source detected with a magnitude $ K_{0} $
at a given position $ x_{i\bullet } $ will be magnified by an additional
factor of $ \geq A $ for a fraction $ \sim 1/A^{2} $ of the time
(Schneider et al. \cite{Sch92}). Hence, continuous monitoring of detected
sources would allow identification of microlensing events. Such events
can be distinguished from variability of the source stars or from variable
patchiness in the dust obscuration, through their achromaticity and their
generic set of possible time profiles. The determination of the microlensing
probability and duration distribution can be used to constrain the density
and velocity distribution of low-mass stars in the GC. We demonstrate this
connection qualitatively by defining a typical timescale for magnification
by more than $ A $ as $ t(>\! A)\equiv \left. \sqrt{\sigma _{\star }(>\! A)}\right/ v_{\bot }\propto \left. \sqrt{m_{\star }}\right/ v_{\bot } $
(Although the actual event duration depends on the complex caustics geometry,
we use this definition as a crude estimate). The projected rms transverse
velocity of the low mass stellar lenses in the GC is (Alexander \cite{Ale99a})
\begin{equation}
\label{eq:vt}
v^{2}_{\bot }\approx 0.37\frac{GM_{\bullet }}{p}\, ,
\end{equation}
 where $ p=\theta R_{0} $ is the projected distance from the BH. Figure
\ref{fig:tscale} compares $ t(>\! A) $ of magnified images of stationary
sources lensed by solar mass stars in the shear field of SgrA$ ^{\star } $
with the timescale of lensing by isolated stars (with the same velocity
field). The duration of a microlensing event strongly depend on the position
of the image, and is increased from several weeks to several months near
$ \theta _{E} $. The formal divergence of $ t(>\! A) $ near $ \theta _{E} $
is truncated in practice by the rapidly decreasing probability of having
a source with an image close to $ \theta _{E} $ (cf Fig. \ref{fig:apdf}b)
and by the motion of the source, which is neglected here. Nevertheless,
a sharp increase of the microlensing timescales is expected near $ \theta _{E} $.
In contrast with timescales of lensing by isolated stars, the timescales
inside $ \theta _{E} $ fall rapidly to zero in the presence of the BH
because the strong de-magnification by the BH (eq. {[}\ref{eq:Abh}{]})
has to be compensated by very high stellar lens magnification. The lensed
lightcurves should show complex structures, unlike the symmetric smooth
lightcurves of an isolated point mass lens. Examples of such lightcurves
can be found in the literature (Mao \& Paczynski \cite{Mao91}; Gould \&
Loeb \cite{Gou92}; Bolatto \& Falco \cite{Bol94}, and in particular Wambsganss
1997).

\section{Discussion and Summary}

\label{sec:discuss}

The presence of a dense stellar cluster around a super-massive BH changes
considerably the lensing properties of both the stars and the BH: the magnification
cross-section, the magnification probability, the variability timescale
and the lightcurve structure. The modification of the microlensing properties
of the stars is most pronounced near the Einstein radius of the BH. Previous
work on lensing by the BH alone (Wardle \& Yusef-Zadeh \cite{War92}; Alexander
\& Sternberg \cite{Ale99b}) focused on stellar sources from the inner
central cluster, for which $ \theta _{E}\sim 0.02'' $. Here we have
considered lensing of distant background stars by the BH and the stars
around it, for which the angular scale is $ \theta _{E}\sim 2'' $.

We applied our calculations to the BH and the dense central cluster in
the GC, and found that the lensing probability is increased over that of
an isolated BH by factors ranging from $ \sim 2 $ for low magnifications
($ \gtrsim 5 $) to $ \sim 3 $ for high magnifications ($ \sim \! 50 $).
We estimate that in any snapshot of the central $ 2'' $ region around
the GC to a $ K $-magnitude limit of $ 23^{\mathrm{m}} $, at least
one distant background star will be magnified by a factor $ \ga 5 $
for $ 10\% $ of the time. The duration of a microlensing event can be
lengthened by an order of magnitude near the Einstein radius of the BH.
Moreover, the gravitational shear of the BH changes considerably the shape
of the microlensing lightcurves so that they are no longer symmetric and
smooth as they are for isolated stars.

Variability provides the distinguishing signature of microlensed stars.
Image subtraction can be used to pick out lensed images of distant background
sources from the many foreground stars. The achromaticity of the lightcurves
and their generic (non-periodic) temporal structures can be used to separate
them from those of intrinsically variable stars. Since the variability
timescale scales as $ \sqrt{m_{\star }}/v_{\bot } $, the lightcurves
contain information on the stellar mass function, the stellar density,
and the stellar velocity dispersion as functions of the projected distance
from the BH. In particular, the lensing timescales could probe mass segregation
deep in the potential of the BH (Miralda-Escud\'{e} \& Gould \cite{Mir00}).
The vast majority of the stellar lenses are low-mass faint stars that are
many magnitudes dimmer than the current detection limit. Microlensing can
probe the statistical properties of these yet unobserved stars.

It is important to note that our quantitative results are subject to a
number of approximations and uncertainties. For the sake of simplicity,
we ignored the fact that the background stars may have a broad distribution
of distances and hence different values for the Einstein radius. In addition,
the model for the surface density and luminosity function of the distant
background stars suffers from large uncertainties given the quality of
present-day data. We did not attempt to deal in detail with the finite
size of the stellar sources, which becomes relevant at high magnifications.
We also ignored the contribution of blended light from the lensing star
to the inferred flux from the source. However, this contribution is likely
to be small for bright sources and low-mass lenses. Lastly, the semi-analytical
formalism for calculating the magnification is not valid very near $ \theta _{E} $
and so our quantitative predictions for $ \left\langle N_{i}(>\! A)\right\rangle  $
are less reliable for very high values of $ A $. More precise calculations,
for example by the ray-shooting method (Wambsganss \cite{Wam97}; Wambsganss
\& Kubas \cite{Wam00}) will be required to improve the accuracy of the
predictions.

Finally, we emphasize that our calculations can be applied to any extragalactic
supermassive black hole. For example, the core of M87 harbors a black hole
mass of $ \sim 2\times 10^{9}M_{\odot } $. The Einstein radius of this
black hole for sources at infinity is $ \sim 1'' $, close to that of
SgrA$ ^{\star } $, and there is no sign for obscuration by dust within
its boundary. However, it is more likely that the sources will be stars
in M87 a few kpc behind the black hole, for which $ \theta _{E}\sim 0.03'' $.
For the higher black hole mass in M87, $ \epsilon \sim 5\times 10^{-10} $
and so microlensing events of stellar sources which are embedded in that
galaxy will have durations comparable to those around SgrA$ ^{\star } $
since $ t\sim \left( \epsilon /M_{\bullet }\right) ^{1/2}\left( D_{l}\theta _{e}\right) ^{3/2} $,
where $ D_{l} $ is the distance from the observer to the BH. Since crowding
and flux sensitivity severely limit the detection of individual stars at
the distance of M87, one may search for microlensing of surface brightness
fluctuations, the so-called pixel lensing (see, e.g. Gould 1996). The Next
Generation Space Telescope (\emph{NGST}; http://ngst.gsfc.nasa.gov/), with
its sub-nJy sensitivity in the wavelength band of $ \sim 1 $--$ 3.5\mu  $m
and its $ \sim 0.06^{\prime \prime } $ resolution, is ideally suited
for such a study.

\acknowledgements

We thank Joachim Wambsganss for communicating results from ray-shooting
calculations prior to publication (Wambsganss \& Kubas 2000) and an anonymous
referee for useful comments. The KLF source model was revised following
results communicated to us by J. Chanam\'e, A. Gould \& J. Miralda-Escud\'e
prior to publication. AL thanks the Institute for Advanced Study for its
kind hospitality during the initiation of this paper. This work was supported
in part by NASA grants NAG 5-7768, 5-7039 and NSF grants AST-0071019, AST-0071019
(for AL).

\clearpage

\end{document}